# Information Governance as a Socio-Technical Process in the Development of Trustworthy Healthcare AI

**Nigel Rees[1], Kelly Holding[1], Mark Sujan[2]**

[1]Welsh Ambulance Service NHS Trust, St Asaph, UK

[2]Human Factors Everywhere Ltd, Woking, UK

**Abstract**

In order to develop trustworthy healthcare artificial intelligence (AI) prospective and ergonomics studies that consider the complexity and reality of real-world applications of AI systems are needed. To achieve this, technology developers and deploying organisations need to form collaborative partnerships. This entails access to healthcare data, which frequently might also include potentially identifiable data such as audio recordings of calls made to an ambulance service call centre. Information Governance (IG) processes have been put in place to govern the use of personal confidential data. However, navigating IG processes in the formative stages of AI development and pre-deployment can be challenging, because the legal basis for data sharing is explicit only for the purpose of delivering patient care, i.e., once a system is put into service.

In this paper we describe our experiences of managing IG for the assurance of healthcare AI, using the example of an out-of-hospital-cardiac-arrest recognition software within the context of the Welsh Ambulance Service. We frame IG as a socio-technical process. IG processes for the development of trustworthy healthcare AI rely on information governance work, which entails dialogue, negotiation, and trade-offs around the legal basis for data sharing, data requirements and data control. Information governance work should start early in the design life cycle and will likely continue throughout. This includes a focus on establishing and building relationships, as well as a focus on organisational readiness deeper understanding of both AI technologies as well as their safety assurance requirements.

## 1    Introduction

The development of data-driven Artificial Intelligence (AI) technologies, such as machine learning (ML), generally consists of three phases involving data capture and pre-processing, model building and validation, and real-world implementation and deployment (Coiera, 2019). In the case of healthcare, AI developers require access to health and care data, which might also include potentially identifiable patient data. Information Governance (IG) processes have been put in place to oversee the use of personal confidential data. However, navigating IG processes in the formative stages of AI development and pre-deployment can be challenging because the mechanisms for data sharing for the purpose of assuring the safety of AI applications are complex and evolving.

The uncertainty about IG processes governing access to health and care data is problematic not least because the development of trustworthy healthcare AI needs to be based on prospective and ergonomics studies that enable iterative and incremental assessment of what happens when AI is introduced into the wider socio-technical system (Sujan, Pool and Salmon, 2022; Vasey *et al.*, 2022). Many studies evaluating healthcare AI are retrospective and focus on the performance of algorithms rather than on the safety and assurance of the service within which the AI is going to be used (Sujan *et al.*, 2019). As a result, the evidence base for the safety and efficacy of these technologies remains weak and is at a high risk of bias (Nagendran *et al.*, 2020; Wu *et al.*, 2021). Often, subsequent prospective evaluation studies demonstrate that one cannot assume that results from retrospective evaluation translate smoothly into successful adoption and deployment in clinical systems (Blomberg *et al.*, 2021; Beede *et al.*, 2020).

Development and retrospective evaluation of healthcare AI are typically performed with a technology-centric focus, with an emphasis on technical issues such as data quality and the potential for bias in the data (Challen *et al.*, 2019). There is a risk that IG is regarded as a deterministic and external process rather than as an integral and formative part of the development life cycle. From a Human Factors and Ergonomics (HF/E) perspective, the development, governance, and deployment of novel and disruptive technologies, such as healthcare AI, should be studied as interacting socio-technical processes rather than as technical and procedural activities in isolation (Sujan *et al.*, 2021).

The contribution of this paper is a reflection on the practical experiences of managing IG processes for the development of trustworthy healthcare AI from a socio-technical systems perspective using the example of an AI system to support the recognition of out of hospital cardiac arrest (OHCA) calls in a Welsh ambulance service clinical contact centre. The next section (Section 2) provides an overview of the current state of IG processes and requirements in Wales. In Section 3 we describe the case study and interpret from a socio-technical systems perspective our experiences of managing IG. Then, in Section 4 we propose recommendations for integrating IG practices into the development life cycle of trustworthy healthcare AI. Concluding remarks are presented in Section 5.

## 2      The Role of Information Governance

Information governance (IG) refers to frameworks and processes aimed at ensuring that information and data are handled in a secure, confidential, and appropriate manner. Organisations processing health and care data need to consider whether they require and meet a legal basis to satisfy data protection legislation. Within the context of the development and deployment of healthcare AI, IG processes are important to ensure data privacy and security, ethical use and appropriate data quality and accuracy.

In Wales, IG processes need to ensure that the requirements of the UK GDPR (General Data Protection Regulation) and the common law duty of confidentiality (CLDC) are met. The UK GDPR applies to personal data, whilst the CLDC applies to confidential patient data. When processing confidential patient information, having a legal basis under the UK GDPR (Article 6 and Article 9) does not remove the need for an appropriate legal basis under the CLDC. The CLDC legal basis that permits data sharing for the purpose of individual (or direct) care is implied consent.

Patients expect their information to be accessed by those treating them, and therefore their consent can be presumed from that expectation. Data sharing for individual care is limited to those within a patient's health and care team, who have a legitimate relationship with that person (and therefore a need to access their information to treat them). Even if a healthcare professional spends all their working hours providing direct care to many people, they only have a legitimate relationship with those individuals for whom they care directly. This limitation should not be seen as a barrier. It should be recognised as an important limit on what sort of data sharing can rely on implied consent as its legal basis. It is a necessary boundary imposed to maintain patient trust in health professionals.

The CLDC applies to confidential patient information. To comply with the CLDC, a CLDC legal basis is required, i.e., implied consent, explicit consent, public interest, required by law, or permitted or approved under a statutory process that sets aside the CLDC, or in the best interests of a patient who lacks capacity. Under Article 5 of the UK GDPR, personal data must be processed lawfully, fairly and in a transparent manner in relation to the data subject. A lawful basis under UK GDPR Article 6 and Article 9 (for special category data including health data) is required, such as performance of a task carried out in the exercise of official authority of the controller. The CLDC and UK GDPR are distinct legal regimes whereby different lawful bases apply and have different requirements.

The boundaries of direct care are often difficult to interpret in practice, and this can cause confusion about whether the purpose for which the information being shared is, in fact, direct care (which has ramifications for the legal basis). Furthermore, Article 5(1)(a) of UK GDPR requires personal data to be processed lawfully. This includes statute and common law obligations, whether criminal or civil, and so processing will be unlawful under UK GDPR if it results in a breach of a duty of confidence, or a breach of the Human Rights Act 1998.

The Confidentiality Advisory Group (CAG) provides authorisation, on behalf of the Secretary of State for Health, to lawfully hold identifiable data on patients without their consent. CAG advises the Health Research Authority (for research) and the Secretary of State (for non-research) on whether there is sufficient justification to use the data. CAG applies to England and Wales. The CAG reviews research and non-research applications and advises whether there is sufficient justification to access the requested confidential patient information. Using CAG advice as a basis for their consideration, the HRA or Secretary of State for Health will take the final approval decision. This provides permission to implement Section 251 of the NHS Act 2006 (originally enacted under Section 60 of the Health and Social Care Act 2001), which allows identifiable patient information to be used without consent in very specific circumstances.

In practice it can be challenging and contentious to navigate the requirements of the UK GDPR and the CLDC along with the boundaries between delivery of direct care and other related crucial activities such as ensuring trustworthiness of AI in use. These practical challenges were highlighted in the high-profile case involving the Royal Free London NHS Foundation Trust and Google DeepMind (Iacobucci, 2017). The Royal Free shared data from 1.6m patient records to enable Google DeepMind to test an app, which can support clinicians in identifying patients at risk of acute kidney disease. The sharing of these data received criticism from the Department of Health's senior advisor on data protection, who expressed concerns about the inappropriate legal basis. Google DeepMind's argument was that the arrangement was covered by the implied consent rule under the common law duty of confidentiality (CLDC), which allows the NHS to use and share data, including with third parties, on the basis of implied consent if it is for the purpose of direct patient care. However, the senior advisor Caldicott said that patients should have been informed because the data were initially used for testing the app which was not a direct care activity. The Royal Free defended their position by arguing that it would not have been possible to sign off the product as clinically safe had it not been tested using real patient information, and that data used to develop it was crucial to demonstrate its safety before being made available for use.

It is tempting to regard the legal requirements set out by the UK GDPR and the CLDC as clear cut and static. As the Royal Free and Google DeepMind case suggests, however, there are many uncertainties and complexities when the legislation is applied to a novel and developing field such as healthcare AI. The legislation leaves room for interpretation, there are many different stakeholders with their own requirements and concerns, and collectively they must navigate the complexities of ensuring confidential, ethical, and appropriate use of data. In the next section we illustrate this socio-technical nature of IG processes through practical experiences from a project concerned with the development of safety assurance (safety case) for an AI application for use in an ambulance service context.

## 3    Information Governance from a Socio-Technical Systems Perspective

### 3.1    AI for OHCA Recognition

Ambulance services are at the forefront of providing unplanned clinical care to patients and are often the first point of contact for patients presenting with a range of urgent and emergency conditions. The Welsh Ambulance Services NHS Trust (WAST) and many other ambulance services internationally have been exploring the use of Artificial Intelligence (AI) in the delivery of emergency medical services across many areas such as improved clinical decision-making, portable diagnostics, communications, and safety monitoring (Rees *et al.*, 2021; Spangler *et al.*, 2019).

In collaboration with academic and industry partners, WAST has been studying prerequisites for the adoption of an AI system to support ambulance service call handlers in the recognition and early detection of out of hospital cardiac arrest (OHCA) (Sujan *et al.*, 2022). OHCA represents one of the most significant challenges for reducing premature deaths, with each minute of delay to defibrillation reducing the probability of survival by about 10% (Deakin, Shewry and Gray, 2014). However, recognition of OHCA is difficult, and the evidence suggests that around 25% of OHCA are not picked up by call centre operators (Blomberg *et al.*, 2019). This OHCA recognition AI system is intended to identify important patterns in live audio from emergency calls in order to prompt clinical contact centre operators to initiate life-saving care earlier. Retrospective studies of the AI system in Denmark and Sweden found that the AI outperformed call handlers in the recognition of OHCA (Blomberg *et al.*, 2019; Byrsell *et al.*, 2021). While these results are encouraging, a prospective evaluation study concluded that the performance of call handlers supported by the AI system did not improve overall (Blomberg *et al.*, 2021). The evaluation study was not designed to explain the reasons for these findings, and further prospective and ergonomics studies are required to understand what happens when AI is introduced into the wider socio-technical system (Sujan *et al.*, 2019; Sujan, Pool and Salmon, 2022).

### 3.2    The ASSIST Study – Safety Assurance of AI for OHCA Recognition in the Welsh Context

The Welsh Ambulance Service was keen to explore the potential adoption of the OHCA recognition AI system. However, the Welsh setting and context differ significantly from the settings where the previous evaluation studies had been undertaken. For example, many areas of Wales are rural with significant travel times whereas the evaluation studies had been undertaken in urban, densely populated areas. In addition, Welsh English differs from other local forms of the English language, and there are also Welsh speakers who require consideration. Hence, the ambulance service required assurance prior to adoption that the AI system was sufficiently trustworthy in the Welsh context.

The ASSIST study aimed to frame the OHCA recognition AI system as part of the wider clinical system of the ambulance service. The objectives of the study were to (1) explore ambulance service stakeholder perceptions on the safety of OHCA AI decision-support in call centres, and (2) to develop a clinical safety case (Sujan *et al.*, 2016; Sujan and Habli, 2021) for the system. Development of the clinical safety case was considered service improvement and received approval by the Medical and Clinical Services Directorate of the Welsh Ambulance Service NHS Trust.

In order to develop this assurance, it was necessary to customise and assess the OHCA recognition AI system with data representative of this Welsh context. The system developer required, therefore, access to data from WAST, in this case access to emergency calls made to the clinical contact centre. These calls are routinely recorded and subsequently audited within WAST, which means the data would have been available. However, IG processes needed to be managed to ensure that data were shared in a secure, confidential, and ethical way, and in accordance with legal requirements.

## 3.3  Negotiating data sharing, data requirements and data control

From the outset, the project team were aware that data needed to be shared and that, consequently, IG processes would need to be followed. However, early on it became apparent that this was not simply a matter of looking up the relevant legislation and guidance. This was not least due to the uncertainty of everyone involved and everyone who was consulted about the specifics and detail of appropriate IG processes when applied to such a novel case. We observed significant hesitation across many levels of stakeholders in the IG process, both within WAST and externally, to go beyond informal guidance and advice and to commit to binding decisions about data sharing arrangements.

We realised, therefore, that IG processes would need to be negotiated and defined as the project unfolded. This can be regarded as a form of "articulation work", or in this case "information governance work", i.e., activities that are required to make something work in practice, but which are often not explicitly recognised and designed as part of the innovation or intervention process (Elish and Watkins, 2020). In this sense, IG becomes very much a socio-technical activity.

### 3.3.1 Data sharing

Clinical contact centres generate data through emergency calls and save these locally. Some of these calls would have needed to be shared with the developer of the AI system in order to customise and test the system. In addition, metadata would have needed to be made available, such as call characterisation, time, date, length of call etc. Following a number of technical meetings between WAST and the technology developer it became clear that full anonymity of data could not be assured due to the volume and the format of the emergency call recordings. Furthermore, voice in itself can be considered personal and potentially identifiable data even if all demographic data (e.g., name, location) has been removed, because voice might be used to identify gender, age, education, language, geographical and socio-cultural origins, and health. Seeking explicit consent from callers to use their data was not considered feasible nor appropriate due to the psychological harm this may cause considering the potentially distressing, sensitive and life-threatening scenarios. Implied consent did not appear to apply because safety assurance is not considered part of direct patient care. However, clinicians with whom we engaged both within WAST and externally suggested that survival from OHCA remains low, and hence there was an ethical duty to use such data for the public good, which may also be the expectation of patients.

These arguments were shared in informal meetings with the Information Commissioner's Office (ICO), Digital Health Care Wales (DHCW) and others in regulatory and advisory roles, but no formal advice was received beyond recognition that this context did not fit easily within established guidelines and principles. The diversity of priorities and perspectives of these stakeholders, and the evolving nature of the field suggest that constant dialogue and negotiation are required to interpret IG processes for a given context. Such negotiations can be drawn out and, in the case of ASSIST, went on beyond 18 months.

### 3.3.2 Data requirements

Data sharing needs to be supported by Data Protection Impact Assessment (DPIA). The DPIA would need to be completed by WAST who hold the data. However, while drafting the DPIA relevant staff at WAST recognised that they did not possess technical details and technical knowledge to fully undertake and populate the DPIA. Collaboration with the technology developer was required to fully determine data requirements, and to understand how these data could be acquired and shared. This required significant amounts of organisational and technical effort and trust, supported by good governance and a suite of documents, including collaborators' agreements and non-disclosure agreements that needed to be in place before technical details could be discussed.

The initial understanding was that the machine learning model did not require patient identifiable data. This was reflected in agreements with the technology developer, which stated that anonymised information would be used. However, in discussions it became clear that further metadata was required and that effective anonymisation of emergency call data would not be feasible (see above). From an organisational perspective, WAST was not sufficiently prepared, nor did it possess suitable technical expertise, to foresee the technical nuances and complexities of the data requirements for the safety assurance of the AI system. These needed to be discovered and subsequently negotiated with the technology developer over the course of several months.

### 3.3.3 Data control

A key requirement within UK GDPR (Article 4) is the identification of data controller and data processor. The data controller is the person or entity determining the purposes and means of the processing of personal data. The data processor is the person or entity processing personal data on behalf of the data controller. Following informal discussions with the ICO and DHCW it was established that there needed to be strong control over the use of data. This would entail uploading the AI software onto WAST systems and giving access to AI engineers from the technology developer to manage the AI system locally at WAST. This arrangement would result in access to data rather than egress of data.

In this arrangement, WAST is the controller of the data who are instructing the technology developer to process data locally within the WAST environment. However, there is the risk that this assumes that safety assurance for the AI system is the responsibility of the user of the AI technology, and it might shift the burden of safety assurance from the AI developer to the deploying organisation. In practice, neither the technology developer nor staff at WAST had prior experiences with this kind of structured safety assurance. Therefore, they needed to rely on input from external safety engineering and human factors experts who required access to both WAST data as well as data from the technology developer.

Furthermore, this surfaced differences in understanding and expectations of how the AI system would operate prior to formal deployment. There were concerns and hesitation on part of the technology developer about the feasibility of creating a local copy of the AI system, without integration into their own development platform and processes. The initial agreements suggested that data would be supplied by WAST, and the technology developer could feed these into their usual customisation and testing processes.

The lack of safety assurance expertise and the need to involve external experts, combined with changes in the processing arrangements for data caused considerable confusion, uncertainty and resulting hesitation among the different parties. These relationships require trust, which needs to be developed over time in dialogue.

## 4 Integrating Information Governance Practices into the Development of Trustworthy Healthcare AI

### 4.1 Information governance work

Information governance processes are critical for ensuring health and care data are used confidentially, ethically, and appropriately. However, the complexities of IG processes can be difficult to manage and navigate because the field of healthcare AI is developing quickly, and new questions and challenges arise. One such challenge is the development of safety assurance for healthcare AI products. This requires collaboration between technology developers, healthcare providers exploring potential adoption, as well as other stakeholders such as regulators and, of course, patients. Within the ASSIST study, most of the 2-year project period was spent on learning about IG and negotiating the legal basis for data sharing, details of data requirements, and appropriate data control. IG processes for the development of trustworthy healthcare AI are not unambiguously defined, nor immediately obvious in their interpretation, and, hence, IG for such purposes might be best understood as a (longer-term) socio-technical process involving dialogue, negotiations, and trade-offs.

This perspective aligns with the broader field of Science and Technology Studies (STS), which suggests that the development, governance, and deployment of novel technologies should be studied as interacting socio-technical processes rather than as technical activities in isolation. For example, adopting an STS perspective, Elish and Watkins explored how the adoption of a deep learning AI system to improve the diagnosis and treatment of sepsis created gaps in the delivery of care and challenged established social structures and hierarchies, which needed to be bridged and repaired (Elish and Watkins, 2020). They identified "repair work", as an instance of articulation work, as an important component of the innovation process. Similarly, Winter and Carusi identified "trust work" as an important part of building trust in AI, which arises from the socio-technical engagements between different stakeholders in the development and validation process of the AI (Winter and Carusi, 2022). In this sense, the experiences reported here around negotiating data sharing, data requirements and data control could be conceptualised as "information governance work". Such information governance work is an essential part of the successful management of IG processes for the development of trustworthy healthcare AI.

### 4.2 Organisational readiness

Through the information governance work, it became clear that all participating organisations (WAST, technology developer and regulatory bodies) would benefit from greater organisational readiness for dealing with the specifics of IG processes for healthcare AI. Organisational readiness refers to the willingness and the ability to adopt a change (Weiner, 2009), in this case related to the adoption of healthcare AI in an ambulance service context. As far as WAST was concerned, the organisation certainly was very willing to explore the adoption of the AI system but lacked significantly in their ability to foresee what was required and what the potential impact might be. The deploying organisation needs to consider their data readiness, including contractual arrangements, processes for managing DPIA, and providing transparent information about AI partnerships to the public. In addition, there needs to be greater technical awareness around AI technologies to enable deploying organisations to engage meaningfully with technology developers.

Questions can also be asked about how familiar AI developers currently are with standards for safety assurance and the underpinning concepts, e.g., around clinical risk management and clinical safety cases. Many developers of AI technology might not come from a medical device background, and they might have little prior experience with the design of health information technology (Habli *et al.*, 2018). There is a need to build capacity and knowledge about safety assurance practices for AI and digital technologies within the health sector (Sujan and Habli, 2021).

Greater organisational readiness underpinned by broader knowledge of AI technologies and their safety assurance is foundational for successful information governance. This enables stakeholders to engage more constructively in dialogue and have a better understanding of how IG requirements might be interpreted in the development of trustworthy healthcare AI.

## 5   Conclusion

We have framed our experiences of navigating the complexities of IG processes from a socio-technical systems perspective. IG processes for the development of trustworthy healthcare AI rely on information governance work, which entails dialogue, negotiation, and trade-offs around the legal basis for data sharing, data requirements and data control. Information governance work should start early in the design life cycle and will likely continue throughout. This includes a focus on establishing and building relationships, as well as a focus on organisational readiness and deeper understanding of both AI technologies as well as their safety assurance requirements.

## 6   Conflict of Interest

The authors declare that the research was conducted in the absence of any commercial or financial relationships that could be construed as a potential conflict of interest.

## 7   Author Contributions

**NR**: Conceptualization, Writing – Original Draft, Funding Acquisition; **KH**: Writing – Review and Editing; **MS**: Conceptualization, Writing – Original Draft, Funding Acquisition.

## 8   Funding

This study received funding from the Assuring Autonomy International Programme, which is a joint initiative between the University of York and Lloyd's Register Foundation.

## 9 Acknowledgments

The authors gratefully acknowledge discussions about information governance with staff at Corti ApS, who developed the out-of-hospital cardiac arrest recognition software.

Edward Harry (WAST), Lauren Williams (WAST), Harold Thimbleby (Thimbleby Works) and Ibrahim Habli (University of York) were part of the ASSIST project team.